\def\={~=~}
\def\+{~+~}
\def\-{~-~}
\def\eq{~\equiv~}
\def\tilde{\widetilde}
\def\beq{\begin{equation}}
\def\eeq{\end{equation}}
\def\beqn{\begin{eqnarray}}
\def\eeqn{\end{eqnarray}}
\newcommand{\calle}[1]{(\ref{#1})}
\newcommand{\CH}{ {\cal H} }
\def\Htilde{{\tilde H}}
\def\Chi{X}

\newcommand{\lb}{\left\lbrack}

\newcommand{\rb}{\right\rbrack}

\newcommand{\bye}{\end{document}}
\newcommand{\ket}[1]{\left| \, {#1} \, \right> }
\newcommand{\beql}[1]{\beq}
\newcommand{\eeql}{\eeq}
\newcommand{\beqnl}[1]{\beqn}
\newcommand{\eeqnl}{\eeqn}
\documentstyle[12pt]{article}
\begin{document}
\rightline{TAUP 2363-96, NIKHEF 96-034, HWS 9617}

\vspace{1cm}

\begin{center}
{\Large\bf 
Shape Invariance in the Calogero and
Calogero-Sutherland Models}
\end{center}

\vspace{5mm}
\begin{center}
C. J. Efthimiou$^\dagger$\\
Department of Physics and Astronomy\\Tel-Aviv University\\
Tel Aviv, 69978 Israel
\end{center}

\begin{center}
Donald Spector$^{\ddagger}$\\
NIKHEF-H\\
   Kruislaan 409\\
   P.O. Box 41882\\
   1009 DB Amsterdam,
   The Netherlands
       \\ and \\
  Department of Physics, Eaton Hall\\
  Hobart and William Smith Colleges\\
  Geneva, NY 14456, USA$^\sharp$
\end{center}

\vspace{2cm}

\begin{abstract}
We show that the Calogero and Calogero-Sutherland models possess
an $N$-body generalization of shape invariance.  We obtain the
operator representation that gives rise to this result, and discuss
the implications of this result, including
the possibility of solving these models using algebraic methods
based on this shape invariance.  Our representation gives us a
natural way to construct supersymmetric generalizations of these models, 
which are interesting both in their
own right and for the insights 
they offer in connection with the exact solubility
of these models.
\end{abstract}
\centerline{December 1996}

\vfill
\hrule
\noindent
$\dagger$ e-mail address: costas@ccsg.tau.ac.il\\
$\ddagger$ e-mail address: spector@hws.edu\\
$\sharp$ Permanent address
\newpage

\section{Introduction}

Exactly soluble models have long provided a foundation for the
study of quantum mechanical systems.  In the case of quantum systems
with one degree of freedom, shape invariance (see \cite{CKS} and
the references therein) has proven to be
perhaps the most compelling technique for proving exact solubility
and for obtaining the exact solution.  In this paper, we demonstrate
that the Calogero \cite{C} and Calogero-Sutherland \cite{S} models
--- two
of the best known $N$-body exactly soluble quantum models
\cite{OP2,Pol1} --- exhibit
an $N$-body generalization of shape invariance, and we use this as
a tool to analyze some features of these models.  We will not be able to 
use this algebraic result to solve completely the models in question, but
we will outline how this might be done, and make
the preliminary steps in that direction.  Clearly, an algebraic
explanation of exact solubility in these models would be a significant
improvement in our understanding of these models, and so our
analysis of the shape invariance structure of these models is valuable.

To lay the appropriate groundwork for our paper, we quickly review
shape invariance with one degree of freedom
in the following section.
Then we turn to 
the heart of the matter,
obtaining a representation for the Calogero and Calogero-Sutherland
Hamiltonians in terms of a set of operators similar to raising and
lowering operators, and we use this representation to demonstrate the
$N$-body shape invariance of these models. 
The power of shape invariance in the one-dimensional case makes clear
the potential significance of this result.
With this in mind,
we then establish a variety of connections, quantitative
and qualitative, between the Calogero and Calogero-Sutherland
models on the one hand, and the shape invariant one-dimensional models
on the other hand, which provides additional evidence that the $N$-body
shape invariance and
one-variable shape invariance are indicators of common
features in these various models.  Our operator representation 
is also valuable as it leads to a natural construction
of the supersymmetric extensions of the $N$-body models we are studying; 
indeed, we
believe this is the first construction of the supersymmetric Calogero-Sutherland
model.  An interesting consequence that we find 
of the $N$-body shape invariance is
that the supersymmetric extension in each of these 
models is not unique.  Note that we examine
these supersymmetric extensions, not simply because they are interesting
in their own right, but because their analysis may well lead to the
solution of the original non-supersymmetric models, an approach we explore
in the penultimate section of this paper.

Readers familiar with shape invariance may be familiar with
the result that, for example, the solution of the hydrogen
atom may be obtained via shape invariance.
Does this not represent the application of shape invariance
to a problem with three degrees of freeedom?  The answer is no; shape
invariance in this case is actually only
 applied to the effective one-dimensional
radial problem.  As another
example, the multidimensional harmonic
oscillator completely factorizes into one-dimensional harmonic
oscillators, and so can be solved by
applying shape invariance for a single degree of freedom repeatedly.
In this paper, we are examining an intrinsically $N$-body property, one that
does not trivially reduce to a one-dimensional problem.

Recently, Ruhl and Turbiner \cite{RT} have explained the integrability of
the Calogero and Calogero-Sutherland models using Turbiner's definition
of exact solubility \cite{Tur}.
In this approach, given an infinite set of finite dimensional
spaces $V_0, V_1, V_2, \dots$, forming an infinite flag
$V_0 \subset V_1 \subset V_2 \subset \dots $, an exactly  soluble
operator $h$  is described as a linear operator that preserves the
infinite flag of the spaces. With this definition and a series
of lengthy calculations, they show that
the Calogero and Calogero-Sutherland models are exactly soluble.
Our goal is something different.  In the same way that shape invariance
has illuminated our understanding of exact solubility in one-dimensional
quantum mechanics (even in those models already known to be
exactly soluble), demonstrating
that an apparent analytic coincidence actually arises from
an elegant and simple algebraic structure,
so too do we expect that the algebraic relationships
we have uncovered will ultimately lead to a ready and efficient 
determination not only of the exact solubility of these models, but indeed
of the solutions themselves.  Thus we are
utlimately interested in the algebraic
understanding of these models that shape invariance can provide,
not simply in the ability to solve the models {\it per se}.

\section{Shape Invariance for Systems  with a Single Degree of Freedom}
\label{sec:2}

Why is the condition of shape invariance signficant?  In the case of systems
with one degree of freedom, shape invariance leads not only to exact
solubility, but indeed to the exact solution.  
In fact, the algebraic methods of shape invariance probably
give the clearest
understanding of why these models are soluble and certainly
give the easiest computational methods for obtaining the energy
eigenvalues and wavefunctions.  We are hopeful that our
$N$-body generalization will be similarly powerful.  In order to consider
how to proceed in the $N$-body case, we here summarize the salient
features of the one-dimensional case.

Consider a Hamiltonian $H$ which is a function of some 
parameter\footnote{$\alpha$ may as  well stand as a collective symbol to
indicate dependence of $H$ on many parameters.} $\alpha$.
One wishes to find the eigenfunctions and eigenstates of $H(\alpha)$.

Suppose one can write $H(\alpha) = A^\dagger(\alpha)A(\alpha)$.  As is
well known from studies of supersymmetry, the partner Hamiltonian
$\Htilde(\alpha)=A(\alpha)A^\dagger(\alpha)$ has essentially the same spectrum.
To be more precise, if
$$
    H\psi = E\psi~,
$$
then 
$$
     \Htilde (A\psi) = A (H\psi) = E (A\psi)~,
$$
and
likewise, if 
$$
    \Htilde \tilde \psi = {\tilde E}\tilde \psi~,
$$
then
$$
    H (A^\dagger \tilde\psi) = {\tilde E}(A^\dagger \tilde\psi)~.
$$
Thus, the spectrum of non-zero energies is the same
for the two Hamiltonians, and the corresponding
 states that have the same energy are easily constructed from each other
by acting with $A$ or $A^\dagger$. 
The only place they may differ in their spectra is that, for example,
$H$ may have a zero-energy eigenstate while $\Htilde$ does not.  
Note too that the representation in terms of $A$ and
$A^\dagger$ ensures that the two Hamiltonians are positive semi-definite,
which means a zero energy state, if it exists, is the ground state.
Furthermore,
such a zero energy eigenstate
is especially easy to construct, as one need only
solve $A\psi = 0$, a first-order rather
than a second-order differential equation.

Shape invariance exists when the two partner
Hamiltonians describe the same
physical model at different values of the coupling constant(s),
i.e., when they are related
in the following way,
\beq
    \Htilde(\alpha) = H(\alpha_1) + R(\alpha_1)~,
\eeq
where $\alpha_1=f(\alpha)$ and $R(\alpha)$ are both $\alpha$-dependent
constants.
When this occurs, one can obtain the entire spectrum of $H$ and
all the wavefunctions once one knows the ground state.  Here is
how it works.

The ground state of $H(\alpha)$ we find by $A(\alpha)\psi_0(\alpha)=0$.
This has zero energy, and hence no ``tilde" partner.  The first
excited state of $H$ is degenerate with the ground state of $\Htilde$.
However,  $\Htilde(\alpha)=H(\alpha_1)+R(\alpha_1)$,
and so its ground state is $\psi_0(\alpha_1)$, with energy
equal to $E_1 = R(\alpha_1)$.
Then the first excited state of $H(\alpha)$ is 
$A^\dagger(\alpha)\psi_0(\alpha_1)$
and also has energy $E_1= R(\alpha_1)$.  Iterating this process, one obtains
the exact energies 
\beq
\label{energies}
  E_k\=\cases{ 0~, &if $k=0~,$\cr
              \sum\limits_{j=1}^k\,R(\alpha_j)~,&if $k>0~.$\cr}
\eeq
and corresponding energy eigenfunctions
\beq
  \psi_n(x;\alpha_0) \=
  A^\dagger(\alpha_0)A^\dagger(\alpha_1)\dots A^\dagger(\alpha_{n-1})
  \psi_0(x;\alpha_n)~,
\eeq
where $\alpha_{j+1}=f(\alpha_j)$.

An example of a shape invariant model  is given by
the trigonometric Rosen-Morse model, with Hamiltonian 
\beq
  H = p^2 + {b(b-a)\over \sin^2(ax)} -b^2
  ~.
\eeq
The parameters of this potential are $\alpha=(b,a)$.
Define 
\beqn
  A &=&  {d\over dx} -b\, \cot(ax) ~,\\
  A^\dagger &=& - {d\over dx} -b\, \cot(ax) ~.
\eeqn
Then $A^\dagger A = H(b,a)$ and 
   $A A^\dagger = H(b_1,a_1) +R(b_1,a_1)$
where $b_1=b+a, ~a_1=a$ and
\beq
   R(b_1,a_1) = (b+a)^2 - b^2~.
\eeq
The spectrum of the model is thus given by
\beq
   E_n= (b+n\, a)^2 -b^2~,
\eeq
where $n$ is a natural number $n=0,1,\dots~.$
It is worth noting that at $b=a$, the Hamiltonian $H$ is that of a free
particle in a box.

\section{$N$-Body Models: Exact Solubility and  Shape Invariance}
\label{sec:3}

The use of shape invariance, or some suitable generalization therof, to
solve systems with multiples degrees of freedom remains an unrealized
goal.  Where shape invariance has been applied to such systems, for
example the 3-dimensional hydrogen atom \cite{KNT},
it has actually been applied
to the effective one-dimensional
radial problem that arises in a sector of fixed angular momentum.

Nonetheless, the idea that there should be an algebraic explanation
for the exact solubility of systems with multiple degrees of freedom
remains compelling.  Indeed, it seems unlikely, given our experience
with ordinary shape invariance, that exact solubility should arise
simply as an analytic accident.

Two of the most studied $N$-body integrable models are the Calogero and
Calogero-Sutherland models.  In this section, we demonstrate that these
models are shape invariant, or, to be more precise, satisfy an algebraic
identity that is the natural $N$-body generalization of shape invariance.
We 
expect that the algebraic results we have found will lead, ultimately, to
a ready derivation of the integrability of these models, and to the efficient
algebraic calculation of the actual wavefunctions and energy levels of these
systems.  What we will do in this paper is to obtain these identities, and then
discuss the analysis of these models that our representation permits.

The generalization of shape invariance we obtain is as 
follows.  In the
models in question, we find we can define a set of operators $A_i$
and $A_i^\dagger$ such that the Hamiltonian can be written
as
\beq
   H(\alpha) = A_1^\dagger A_1 + \cdots +  A_N^\dagger A_N~.
\eeq
However, the operators $A_i$ and $A_j^\dagger$ do not commute.
Thus, this identity provides an elegant representation for the Hamiltonian,
but nonetheless one that generalizes, but falls short of, separability.

The compelling result, however, is the following algebraic structure
which turns out to be present in these
models.  If we interchange the $A_i$'s and
$A_i^\dagger$'s, we obtain the associated Hamiltonian
\beq
   {\tilde H}(\alpha) = A_1 A_1^\dagger+ \cdots+ A_N A_N^\dagger~.
\eeq
We are able to identify these models as shape invariant because,
it turns out that 
\beq
  {\tilde H}(\alpha) = H(\alpha_1) + R(a_1)~.
\eeq
This identity generalizes the notion of shape invariance familiar from
models with one degree of freedom.
We conjecture that the ability to write the Hamiltonian in these
two ways underlies the exact solublility of these models.  

In the rest of this
section, we will present the specific forms of these representations.
Then, in this and later sections, we will discuss the applications
of this representation of the Hamiltonians, providing support for
our conjecture, keeping an eye toward how this $N$-body shape
invariance should be
related to the exact solubility of the models in question.
Also, in appendix \ref{appendixA}, we present a general definition
and treatment
of ``shape invariant" potentials for $N$-body potentials in an
analytical framework.

\subsection{The Calogero Model}

The Calogero model is an $N$-body non-relativistic quantum mechanical
model given by the Hamiltonian
\beq
\label{eq:hC}
  H_C = -\sum_i \partial_i^2  +
  \sum_i \sum_j{}'\, {g/2\over (x_i - x_j)^2}~,
\eeq
where the prime in the summation sumbol means `restricted sum',
i.e. summation over the dummy variable $j$ excluding the
value $i$ for which the denominator vanishes.
We can obtain a useful representation of this Hamiltonian as follows.

Let us define the prepotential\footnote{The quantity we term the {\it 
prepotential} is sometimes referred to as the ``superpotential'' in the
literature.  We prefer to use the term ``prepotential,'' 
reserving ``superpotential"
for the quantity whose gradient gives $W_i$, in order to be consistent
with the terminology of supersymmetric field theory.}
\beq
  W_i = \sum_j{}' \, {-\alpha \over x_i - x_j}~~,
\eeq
in terms of which we define the
operators
\beq
   A_i = \partial_i + W_i~,~~~~~
   A_i^\dagger = -\partial_i + W_i~.
\eeq
These operators obey the following algebra:
\beqn
\label{eq:C1}
    \lb A_i,A_j \rb &=& 0~, ~~~
   \lb A_i^\dagger ,A_j^\dagger \rb  = 0~, \\
\label{eq:C2}
   \lb A_i^\dagger ,A_j \rb  &=& -2\, \partial_i W_j   = 
   \cases{ \sum_k{}' {-2\alpha\over (x_i-x_k)^2}~, & if $~i=j~,$\cr   
            {2\alpha\over (x_i-x_j)^2}~, & if $~i\ne j~.$\cr}
\eeqn
Two further relevant identities are that
$$\sum_i W_i = 0$$
and that the curvature associated with the prepotential vanishes,
i.e.,
\beq
\partial_i W_j = \partial_j W_i~~.
\eeq

The first important observation is that the Hamiltonian can be 
written in a simple
way using the $A_i$'s and $A_i^\dagger$'s.
In particular, one finds that
\beq
   H_C(\alpha) = \sum_i A_i^\dagger(\alpha)A_i(\alpha)
\eeq
at coupling constant 
$$
   g = 2\,\alpha\, (\alpha-1)~.
$$
In order to verify this result, note that
\beq
   A_i^\dagger A_i = -\partial_i^2 + W_i^2 -\partial_iW_i ~.
\eeq
To see that this expression yields the Calogero Hamiltonian, it is
necessary to verify that the three-body interactions contained in the $W^2$
term cancel.  (No other term contributes to a direct three-body potential.)
To see this, we consider all terms in $A_i^\dagger A_i$ that involve
$x_i$, $x_j$, and $x_k$, where these three coordinates are
distinct.  The terms are
$$
   {1\over (x_i-x_j)(x_i-x_k)}+{1\over (x_j-x_i)(x_j-x_k)}
  +{1\over (x_k-x_i)(x_k-x_j)}~,
$$
which it is easy to verify is zero by putting everything over a
common denominator.

This representation for the Hamiltonian  in terms of $A_i$'s and
$A_i^\dagger$'s is potentially rather powerful.  Unfortunately,
these operators do not diagonalize the Hamiltonian, due to the 
non-commutation contained in \calle{eq:C2}. 
Nonetheless, the vanishing of the three-body interactions, the
representation of the Hamiltonian as a sum of quadratic terms, and the
pairwise commutation of the $A_i$'s (respectively, the
$A_i^\dagger$'s) does at least give us a structure akin to 
separation of variables.  
It is the rare Hamiltonian that takes this form.

There is however a more significant algebraic identity at work
here.  If we switch $A_i$ and $A_i^\dagger$, we obtain
\beq
  \tilde H_C(\alpha) =
  \sum_i \, A_i(\alpha)A_i^\dagger(\alpha) = H_C(\alpha-1)~.
\eeq
In other words,
\beq
  \sum_i\, A_i(\alpha)A_i^\dagger(\alpha) =
  \sum_i\, A_i^\dagger(\alpha+1) A_i(\alpha+1)~.
\eeq
Interchange of the $A_i$'s and $A_i^\dagger$'s has the same effect as shifting
the coupling constant!  This is the hallmark of shape invariance, and in the
later sections of this paper, we consider the possible connection of this
identity
to the exact solubility of the model.  
We have thus have demonstrated that the Calogero model exhibits
an $N$-body generalization of shape invariance, the first time it
has been found in such a model.  

These operators can be used to represent physically relevant quantities.
An elementary but useful result is that the total momentum can be
written as
\beq
   P_{TOT} = -i\sum_i A_i = i\sum_i A_i^\dagger~,
\eeq
since $\sum_i W_i = 0$.  The quantity $P_{TOT}$ commutes with each of 
the $A_i$ and $A_i^\dagger$ individually, and hence we can classify solutions
in terms of their total momentum
(not surprisingly) in a way that does not interfere
with our operator representation of the Hamiltonian.

The Calogero model has an important variation which we now describe.
One can introduce additional pairwise interactions among the particles
quadratic in the coordinates:
\beq
\label{eq:HhC}
   H_{hC} = -\sum_i \partial_i^2  
  + \sum_i \sum_j{}'\, {g/2\over (x_i - x_j)^2}~,
  + \sum_i \sum_j{}'\, {1\over 4} \, \omega^2\, (x_i - x_j)^2
  + c~.
\eeq
In order to distinguish this from the model we discussed initially, 
we shall coin the term ``harmonic-Calogero model" to describe
the theory with this quadratic interaction.  The discussion 
presented above in the case of the ordinary Calogero model
can be repeated nearly verbatim, with only minor changes.
For simplicity, we here highlight only the important details.

The operators we must define are 
\beqn
   A_i  &=& \partial_i  
   -\sum_j{}' \, {\alpha \over x_i - x_j}   
   +\beta\, \sum_j{}' \,  (x_i - x_j)   ~, \\
   A_i^\dagger &=& -\partial_i 
   -\sum_j{}' \, {\alpha \over x_i - x_j} 
   +\beta\, \sum_j{}' \,  (x_i - x_j)   ~, 
\eeqn
where
\beq
  2\,\alpha\, (\alpha-1)=g~, ~~~~~
  \beta ={\omega\over2\sqrt{N}}~.
\eeq
With these operators, we have both the generalizations of separation
of variables and of shape invariance that we found in the ordinary
Calogero model.  To be precise,
\beq
  H_{hC}(\alpha) = \sum_i \, A_i^\dagger(\alpha) A_i(\alpha) ~,
\eeq
and the constant $c$ in equation \calle{eq:HhC} is
\beq
   c=  -{\omega\over \sqrt{2}}\, \sqrt{N}\, (N-1)\, (\alpha N+1)~.
\eeq
Then
\beq
  \sum_i \,A_i(\alpha)A_i^\dagger(\alpha) =
  \sum_i\, A_i^\dagger(\alpha_1)A_i(\alpha_1) + R(\alpha_1)~,
\eeq
where
\beq
   \alpha_1=\alpha+1~,
   ~~~~~~
    R(\alpha_1) = 
     {\omega\over \sqrt{2}}\, \sqrt{N}\, (N-1)\, 
     (\alpha_1 -
     \alpha)\, N~.
\eeq
Further analysis of this model can be performed exactly as for the ordinary
Calogero model; we leave 
the particular calculations to the interested reader.

\subsection{The Calogero-Sutherland Model}

Our formulation of the Calogero-Sutherland model, and the operators
we use to represent the Hamiltonian, can be developed very
similarly to the Calogero model.  The Calogero-Sutherland
model is given by an $N$-body, non-relativistic quantum mechanical
Hamiltonian,
\beq
    H_{CS}(g) = -\sum_i \partial_i^2 
   +\sum_i \sum_j{}' {g/2\over \sin^2\lbrack a(x_i-x_j)\rbrack }
   +{\rm c}~.
\eeq
In the following, we shall set $a=1$.
We now define a prepotential
\beq
W_i = - \alpha \sum_j{}' \cot(x_i - x_j)~.
\eeq
This leads us to introduce the operators
\beq
  A_i = \partial_i + W_i~,~~~~~
  A_i^\dagger = -\partial_i + W_i~.
\eeq
These operators satisfy a variety of algebraic relations, and are useful
in analyzing the Calogero-Sutherland model.
For these new $A_i$ and $A_i^\dagger$ operators,
we have
\beqn
   \lb A_i,A_j \rb &=&0~, ~~~~~ \lb A_i^\dagger,A_j^\dagger\rb=0~, \\
   \lb A_i^\dagger,A_j\rb &=& -2\, \partial_i W_j = 
    \cases{ \sum_k{}' {-2\alpha\over\sin^2(x_i-x_k)}~, & if $~i=j~,$ \cr
            {2\alpha\over\sin^2(x_i-x_j)}~, & if $~i\ne j~.$ \cr }
\eeqn
It is worth noting again here, as in the previous model, that 
$$
   \sum_i W_i = 0
$$
and that
$$
   \partial_i W_j = \partial_j W_i~.
$$
Setting 
$$
     g = 2\,\alpha\, (\alpha - 1)
$$
 and
\beq
   {\rm c} = - \alpha^2\, {N(N^2-1)\over 3}~,
\eeq
the Hamiltonian can be written as
\beq
    H_{CS}(\alpha) = \sum_i A_i^\dagger(\alpha) A_i(\alpha)~,
\eeq

Just as with the
Calogero model, the three-body interactions automatically cancel.
As before, the only source of three-body interactions is the $W^2$ contribution
to $A_i^\dagger A_i$.  Let us consider all the terms that include $x_i$,
$x_j$, and $x_k$.  For simplicity, let $a$, $b$, and $c$ represent the
differences $x_i - x_j$, $x_j - x_k$, and $x_k - x_i$, respectively.
Then we have terms
$$
  -\cot a \cot b - \cot b \cot c - \cot c \cot a ~.
$$
Now, since $a+b+c=0$, we see
\beqn
          \cot a \, \cot b + \cot a\, \cot c +
          \cot b \, \cot c \= 1~.
\eeqn
Thus we can write the Hamiltonian in terms of the $A_i$'s 
and $A_i^\dagger$'s in a form that generalizes separation of
variables.

And, again as before, there is more.  If we switch the hermitian
conjugates, we get the partner Hamiltonian
\beq
\tilde H_{CS}(\alpha) = \sum_i\, A_i(\alpha) A_i^\dagger(\alpha)~.
\eeq
Remarkably, as with the previous model, we see that interchanging
the hermitian conjugates is equivalent to shifting the coupling constant!
To be precise, we find that
\beq
\tilde H_{CS} (\alpha) =
     H_{CS}(\alpha_1) + R(\alpha_1)~,
\eeq
with 
\beq
  \alpha_1 = \alpha+1~, ~~~~~
  R(\alpha+1) = (\alpha_1^2-\alpha^2) \, {N(N^2-1)\over 3}~.
\eeq
Put it another way,
\beq
      \sum_i\, A_i(\alpha)A_i^\dagger(\alpha) 
      = \sum_i \, A_i^\dagger(\alpha_1) A_i(\alpha_1) 
      + R(\alpha_1)~.
\eeq
We have therefore obtained a generalized shape invariance 
in another $N$-body model.

Note, too, that we again find 
$$
    P_{TOT} = -i\sum_i A_i = +i\sum_i A_i^\dagger~,
$$
where of course $P_{TOT}$ commutes with all the individual $A_i$'s
and $A_i^\dagger$'s..

Notice that in both these models, the key to shape invariance is
that, up to an additive constant, $\partial_i W_i$ and $W_i W_i$
have the same functional dependence, which is the same as that of the
potential.   This
relationship, we expect, and the generalized
shape invariance it implies, truly underlie the solubility of the system.

\section{Generalizations of $1$-body Shape Invariance}

The Calogero and Calogero-Sutherland models share a number of features
in common with familiar shape invariant models of one degree of
freedom.  These similarities go beyond the algebraic identities above,
and relate to the features of the model.  We review here some of
these similarities, laying the groundowrk for a future
solution of these models by means of shape invariance.

\begin{enumerate}
\item
   Interchanging $A_i$ and $A_i^\dagger$ has the same effect on the
   Hamiltonian as shifting the coupling constant.
   This is the fundamental
   hallmark of shape invariance in the one-body problem, and arises as
   well in the $N$-body Calogero and Calogero-Sutherland models.
\item
   These $N$-body models are known to be exactly
   soluble.   Likewise, the shape invariant models with one degree
   of freedom are known to be exactly soluble (indeed, were generally
   known to be so before shape invariance was discovered).
\item
   The $N$-body models in question are the 
   obvious multiparticle generalizations of
   known shape invariant models with one degree of freedom.  More
   precisely, the $N$-body models have potentials which involve only
   pairwise interactions; the pairwise potential in each case is given
   in terms of a single relative coordinate; and this two-body potential
   is just one of the shape invariant, exactly soluble potentials from ordinary
   quantum mechanics with one degree of freedom.
\item
    The partition function computed semiclassically is exact in
    these theories. The evidence for that has been presented in 
    \cite{FM}.
    This is also a feature exhibited by
    shape invariant models with one degree of freedom \cite{CKS}.
\item
    For the case $N=2$, i.e., the two-body problem, we actually directly recover
    the conventional case of shape invariance with one
    degree of freedom, once we
    separate out the total momentum $P_{TOT}$.
\item
    The ground state is easy to construct, by requiring it to be
    a state annihilated by all the $A_i$'s.  This gives us a zero energy
    state, but not for the partner Hamiltonian.  Shape invariance, however
    lets us get the ground state for the partner.
\end{enumerate}

It is worth turning our attention to the last two of these items in particular.
We turn first to construction of the ground state.

To find the Calogero or Calogero-Sutherland ground
state, we use the representation
of the Hamiltonian as $H = \sum_i A_i^\dagger A_i$.
 Since this expression is positive
semi-definite, the states of the theory must have non-negative energy.
Therefore,
if we can find a zero-energy state, this will of necessity be the ground state.

In fact, since the Hamiltonian is the sum of $N$ postive semi-definite terms,
in order for a state $\Phi_0$ to have zero energy,
it must satisfy
$$
    A_i \Phi_0 = 0~,~~~~~\forall ~i~.
$$
It is easy enough to solve this, obtaining
$$
    \Phi_0(x_1,\ldots,x_n) = \phi_0 \, e^{ -\sum_i \int^{x_i} W_i dx_i}~.
$$
As long as this function is normalizable,
it yields a sensible state, and then this
is the ground state wavefunction of this system.

Notice that normalizability forces us to recognize that if there are
solutions to the equation $A_i(\alpha)\Phi_0(\alpha)=0$,
 there are no corresponding solutions
$\tilde \Phi_0(\alpha)$
to $A_i^\dagger(\alpha)\tilde\Phi_0(\alpha) = 0$.  Nonetheless, we can
find the ground state $\Phi^\prime(\alpha)$
of $\sum_i A_i(\alpha)A_i^\dagger(\alpha)$ by exploiting the shape
invariance of the model.  In particular, since
$\sum_i A_i(\alpha) A_i^\dagger (\alpha) 
 = \sum_i  A_i^\dagger(\alpha+1) A_i (\alpha+1)+R(\alpha+1)$,
the ground state $\tilde\Phi^\prime(\alpha) \propto \Phi(\alpha+1)$.
The energy is given by the constant $R(\alpha+1)$.
  Thus shape invariance
allows us immediately to obtain the ground state
energy of $\sum_i A_i A_i^\dagger$, despite the absence of a physical
solution to $A_i^\dagger\tilde\Phi_0 = 0$.

The second point we wish to turn to is the special situation when
$N=2$, i.e., when we have a 2-body Hamiltonian.
In our two models in this case, then, the Hamiltonian is
\beq
   H^{(2)} = A_1^\dagger A_1 + A_2^\dagger A_2~,
\eeq  
where the superscript indicates the number of particles.
We can write the 
Hamiltonian in this case as
\beq
   H^{(2)}  = {1\over 2}(A_1^\dagger + A_2^\dagger)(A_1+A_2)
            + {1\over 2}(A_1^\dagger - A_2^\dagger)(A_1 - A_2)~.
\eeq
Recall that the total momentum
\beq
    P_{TOT} = -i(A_1+A_2)=i(A_1^\dagger+A_2^\dagger)
\eeq
and that both $A_1-A_2$ and $A_1^\dagger-A_2^\dagger$ commute with
$P_{TOT}$.  Consequently, if we define (for simplicty of notation)
$B=A_1-A_2$, we can write the Hamiltonian as
\beq
   H^{(2)} = {1\over 2}\, P_{TOT}^2 +
               {1\over 2}\, B^\dagger B~.
\eeq
Since $[P_{TOT},B]=[P_{TOT},B^\dagger]=0$, the solutions of this model
are
$$
   \psi(x_1,x_2) = e^{ik(x_1+x_2)}\psi_B(x_1-x_2)~,
$$
where $\psi_B(x_1-x_2)$ is an eigenstate of $B^\dagger B$.

What of the partner Hamiltonian,
\beq
{\tilde H}_2 = A_1 A_1^\dagger+ A_2 A_2^\dagger ~~?
\eeq  
Using the same definitions as above, we see that
\beq
{\tilde H}_2 = {1\over 2}P_{TOT}^2 + B B^\dagger
\eeq
Thus, other than the overall center-of-momentum term
$\exp[ik(x_1+x_2)]$ in the wavefunction, solving this problem
simply involves finding the wavefunctions of $B^\dagger B$ and
$B B^\dagger$.  This can be done using the familiar result for shape
invariance for systems with one degree of freedom, and thus we see
that our identity in the $N$-body case does indeed reduce to the familiar
shape invariance, with all its attendant usefulness for exact solubility,
in the two-body case.  Later in the paper we explore some ideas of
how our identity might produce an exact solution in the $N$-body case.

To close this section, we note that there is a generalization of our
separation in the two-body case into $P_{TOT}$ and $B$.  While
unfortunately this generalization does not lead directly to the exact
solution of the problem, it does give a representation of the Hamiltonian
in an alternative basis that is superior in certain respects to the
representation in terms of the $A_i$'s and $A_i^\dagger$'s above.

Let us go to a Jacobi\footnote{Our   
terminology comes from the similarity
of the above transformation to the choice of
Jacobi coordinates.  
In classical mechanics, the Jacobi coordinates
\beqn
y_1 &=& x_1 - x_2
  ~, \\
y_2 &=& {m_1 \, x_1 + m_2\, x_2 \over m_1+m_2} -  x_3
  ~, \\
\cdots &\cdots&  \cdots
  \\
y_{N-1} &=& 
   {m_1\, x_1 + m_2 \, x_2 + \cdots + m_{N-1}\, x_{N-1} \over
   m_1+m_2+\cdots + m_{N-1}} - x_N
 ~, \\
Y &=& {m_1\, x_1  + m_2\, x_2 + \cdots + m_N\,x_N \over m_1+m_2+\cdots+m_N}~,
\eeqn
are used to separate out the center-of-mass motion.} 
basis.
We define
\beqn
B_1 &=& {1\over \sqrt{2}}(A_1 - A_2)
  ~, \\
B_2 &=& {1\over\sqrt{6}}(A_1 + A_2 - 2 A_3)
  ~, \\
\cdots &\cdots&  \cdots
  \\
B_{N-1} &=& {1\over \sqrt{N(N-1)}}
   \lb A_1 + A_2 + \cdots + A_{N-1} - (N-1)A_N \rb
  ~, \\
B_N &=& {1\over \sqrt{N}}(A_1 + A_2 + \cdots + A_N)
 ~.
\eeqn

In this operator basis, we still have the
generalized separability result,
$$
  H = \sum_i B_i^\dagger B_i
$$
and we can still see directly the $N$-body shape invariance result
$$
 \sum_i\, B_i B_i^\dagger (\alpha) =
 \sum_i\,  B_i^\dagger B_i(\alpha_1) + R(\alpha_1)~.
$$
The advantage of this representation is that the commutation 
relations are, though less symmetric, in one respect, a little simpler.
We again have
$$
   [B_i,B_j] = [B_i^\dagger,B_j^\dagger] = 0~,
$$
for all $i$ and $j$.  But now the operators $B_N$ and $B_N^\dagger$
are proportional to the total momentum, and so we have
$$
  [B_N,B_j^\dagger] = [B_N^\dagger,B_j]=0~,
$$
for all $j$.  In addition, this choice automatically pulls out the 
total momentum, while preserving a raising/lowering operator
formalism.  In other words, we can write
$$ 
   H={1\over 2}P_{TOT}^2 + \sum_{i=1}^{N-1} B_i^\dagger B_i
$$
and
$$
  {\tilde H} = {1\over 2}P_{TOT}^2 + \sum_{i=1}^{N-1} B_i B_i^\dagger~.
$$

\section{Supersymmetric Construction and Generalizations}

For the case of shape invariance with one degree of freedom, the 
algebraic results are best understood in the context of supersymmetry. 
Thus we anticipate that the supersymmetric generalizations of the
Calogero and Calogero-Sutherland models offer the best possiblity 
for understanding the exact solubility of these models.  To this end,
we here review the relevance of supersymmetry to shape invariance with
one degree of freedom, and then proceed to construct the supersymmetric
Calogero and Calogero-Sutherland models.  We will see that our representation
of the Hamiltonians in these cases leads to an easy and natural construction
of supersymmetric generalizations for these models.   In fact, we believe that
this is the first appearance of the supersymmetric Calogero-Sutherland model
in the literature; the supersymmetric Calogero model has been discussed
in \cite{FM,BHKV}. Incidentally, we mention that recently
higher dimensional supersymmetric quantum mechanics has been
studied by other authors too \cite{DP}, although with a
different perspective.

Let us first therefore consider quantum mechanics with one degree of freedom.
The degeneracy of the spectra of the two Hamiltonians $H$ and $\tilde H$ can
be understood as a consequence of supersymmetry. One constructs
a single supersymmetric system, with Hamiltonian
\beq
 \CH = \lb\matrix{ H& 0\cr 0&\Htilde\cr}\rb~.
\eeq
Defining the operators
\beqn
  Q &=& \lb\matrix{ 0& 0\cr A &  0\cr}\rb~,\\
  Q^\dagger &=& \lb\matrix{ 0& A^\dagger\cr 0 &  0\cr}\rb~,
\eeqn
we can easily verify  the algebra
\beqn
  [ \CH, Q ] &=& [  \CH, Q^\dagger ] =0~,\\
  \{ Q , Q \} &=& \{ Q^\dagger,  Q^\dagger \} =0~,\\
  \{ Q , Q^\dagger \} &=& \CH~.
\eeqn
The ground state, if it has zero energy, is a supersymmetry singlet;
other than that all states must be paired.

Recognizing that $\sigma_3$ measures
fermion number in this example, one sees that
using shape invariance and
supersymmetry to link together wavefunctions at different fermion number
is what enables one to solve the theory.  
Motivated by this one-body result, we now construct
the supersymmetric
generalization of the $N$-body
Calogero and Calogero-Sutherland models, and then
study the ways in which the wavefunctions and energy levels at different
fermion number are related to each other.  We will find that
going from fermion number $0$ to the maximal fermion number $N$,
one travels from the original Calogero or Calogero-Sutherland 
Hamiltonian to its shape invariance partner.  While we will
not be able to relate these two fermion number levels directly,
we will identify an approach for obtaining such a relationship,
and hence solving the original models via shape invariance.

\subsection{Supersymmetric Generalizations}

One advantageous feature of our representation of the Hamiltonian
and related physical quantities in terms of $A_i$'s and
$A_i^\dagger$'s 
is that it
makes construction of the supersymmetric generalizations
of the Calogero and Calogero-Sutherland models completely
straightforward.  We thus can
and do use the same formalism for both models.

Let us define Grassmann variables $\psi_i$ and $\psi_i^\dagger$
with the anticommutation relations
\beqn
   \{ \psi_i \, , \, \psi_j \} =
   \{ \psi_i^\dagger \, , \, \psi_j^\dagger \} &=& 0~, \\
   \{ \psi_i \, , \, \psi_j^\dagger \} &=& \delta_{ij}~.
\eeqn
We define the supercharges 
$$
  Q = \sum_i\, A_i\,  \psi_i~, ~~~~~
  Q^\dagger = \sum_i\, A_i^\dagger\, \psi_i^\dagger~,
$$
and the Hamiltonian
$$
   H_{susy}=\{ Q^\dagger,Q \}~.
$$
  It is easy to verify
that $ Q^2=(Q^\dagger)^2=0$.
Thus this theory has an $N=2$ supersymmetry; both the real
and imaginary parts of $Q$ square to the Hamiltonian.

The structure of these supersymmetric theories is related to
the generalized shape invariance.  We note that the superalgebra
determines that all states are annihilated by $Q$ or $Q^\dagger$.
The zero energy states are annihilated by both.  Of the remaining
states, those eigenstates of the Hamiltonian annihilated by $Q$
are also in the range of $Q$; and those annihilated by $Q^\dagger$
are in the range of $Q^\dagger$.

\subsection{Relating States at Different Fermion Number}
Ignoring the zero energy states for simplicity, we consider dividing
the space of eigenstates into states of definite fermion number
(which can be anything from $0$ to $N$) and also according to whether
they reside in $\ker~Q$ or $\ker~Q^\dagger$.

\begin{enumerate}
\item
The 0-fermion states:  These states are automatically in $\ker~Q$.
Furthermore, they satisfy the equation 
\beq
   \sum_i\, A_i(\alpha) A_i^\dagger(\alpha) \phi = E\phi~.
\eeq
In other words, the states of the ordinary bosonic Calogero or
Calogero-Sutherland model are the 0-fermion states of the
supersymmetric version.

\item
The $N$-fermion states: These states are automatically in $\ker~Q^\dagger$.
Futhermore, they satisfy the equation
\beq
   A_i^\dagger(\alpha) A_i(\alpha) \phi = E\phi~.
\eeq
In other words, the $N$-fermion states are {\it also} the states of
the ordinary bosonic Calogero or Calogero-Sutherland
model --- except
at a shifted value of the coupling constant, due
to shape invariance.

\item The 1-fermion states in $\ker~Q^\dagger$: These
are obtained by applying $Q^\dagger$ to the zero
fermion states.  Thus, these states are degenerate with and derivable
from the 0-fermion states, i.e., from the original
Calogero or Calogero-Sutherland model.

\item The 1-fermion states in $\ker ~Q$:  The best way to understand these
is as follows.  Let us write them as 
$$  
    \sum_i\, \phi_i(x_1,\ldots,x_N) \psi_i \ket{\emptyset}~.
$$
Then it is straightforward to show that $A_i \phi_i = 0$,
and that 
$$
    \sum_i\, (A_i A_i^\dagger \phi_j - A_i A_j^\dagger \phi_i)
     = \lambda \phi_j~.
$$
Let us define $\Phi = \sum_i \phi_i$.  
One immediately determines that
since $\sum_i A^\dagger_j \propto P_{TOT}$ commutes with the $A_i$,
$$
   A_i A_i^\dagger \Phi = \lambda \Phi~~.
$$
This means that $\Phi = \sum_i \phi_i$ is either a 0-fermion
wavefunction (i.e., a solution of the original Calogero or
Calogero-Sutherland model) or is itself vanishing.  Thus, as long as the sum
does not vanish, the energy of such a wavefunction is degenerate
with a state of the zero-fermion spectrum.

\item The 2-fermion states: Those in $\ker~Q^\dagger$ we obtain from the
1-fermion states in $\ker~Q$. The others we construct similarly
to what we did above.

\item $(N-1)$-fermion states in $\ker~Q$:  These are simply
obtained by applying $Q$ to the $N$-fermion states, and so we
get a spectrum and set of eigenfunctions directly from the
ordinary Calogero or Calogero-Sutherland model at shifted
coupling constant.

\item $(N-1)$-fermion states in $\ker~Q^\dagger$: We can proceed much as we did for the
1-fermion states in $\ker~Q$.  We find that the
$(N-1)$-fermion states in $\ker~Q^\dagger$ can be
written as
$$
   \sum_i\, \chi_i \psi_i \ket{N}~,
$$
where $\ket{N}$ is the state with $N$ fermions in the fermionic subspace
(i.e., proportional to $\psi_1\cdots\psi_N\ket{\emptyset}$),
 and $\chi(x_1,\ldots,x_n)$
is the bosonic piece of the wavefunction.  Then
$$
   A_i^\dagger \chi_i=0
$$
and one finds too that
$\Chi = \sum_i \chi_i$ either vanishes or is itself an eigenstate
of $\sum_i A_i^\dagger A_i$, and so the energy spectrum includes the energy
eigenvalues of the bosonic Calogero or Calogero-Sutherland model
at shifted coupling constant.

\end{enumerate}

\subsection{Comments}

We see from this method that we can related the spectra at different
fermion number.  Ultimately, we expect that refinements in this
procedure will enable one to relate
the spectra of the 0-fermion and $N$-fermion states, which means
relating the Hamiltonians $H$ and $\tilde H$, which in
turn means relating the Hamiltonians $H(\alpha)$ and
$H(\alpha_1)$.  Consequently, we would expect to be able
to solve the model algebraically, just as one does in shape
invariant systems with one degree of freedom.

A brief look at the case $N=2$ is worthwhile.  While we
discussed the case $N=2$ without supersymmetry
and saw how to use generalized shape invariance
to solve such a system, we see here
how supersymmetry organizes the states usefully.  In the
case $N=2$, the 1-fermion states
in $\ker~Q$ for example
are degenerate with and
can be obtained by acting
on the 2-fermion states with $Q$; but they may also
be related to the 0-fermion states as described above.
The bosonic part of the state has components $\phi_1$ and
$\phi_2$, and as long as these do not cancel additively,
the sum $\phi_1 + \phi_2$ is
a 0-fermion state that is degenerate with the 1-fermion
state, and hence in turn with a 2-fermion state.

The case of $N=3$ also deserves special study, as everything
is still very tightly constrained. Here, the spectra of
the 1-fermion states in $\ker~Q$ and the 2-fermion states
in $\ker~Q^\dagger$ are degenerate; indeed, using the labels $\phi_i$
and $\chi_i$ as above, one finds
\beq
   \phi_i = \pm \,c \sum_{i,j}\, \epsilon_{ijk} A_j \chi_k
\eeq
and a conjugate expression for $\chi_i$.  Since these spectra
are degenerate, and in turn are related, respectively, to
$H(\alpha)$ and $H(\alpha_1)$ as discussed in items 4 and 7
in the preceding subsection, one has nearly all the ingredients
in place to relate directly the spectra of 
$H(\alpha)$ and $H(\alpha_1)$ in the $N=3$ case.

\subsection{Non-Uniqueness of Supersymmetric Extensions}

Exactly because the Calogero and Calogero-Sutherland models
are shape invariant, their supersymmetric extensions are not
unique.  This is easy to see, and important to recognize.

Let us consider the pure Hamiltonian
$$
  H= -\partial_i^2 +g\,V +c ~,
$$
where\footnote{In this subsection, we have separated the
coupling constant dependence from the potential.}
$g=2\,\alpha\,(\alpha-1)$.
We can write this Hamiltonian in two distinct ways,
$$
   H= \sum_i \, A^\dagger_i(\alpha)A_i(\alpha)
$$
and
$$
  H=\sum_i \, A_i(\alpha+1)A_i^\dagger(\alpha+1)~.
$$
Up to an additive constant, these Hamiltonians are the same, and thus
describe exactly the same physics.
We can take either one of these representations as the starting
point for our supersymmetric construction.  However,
although both these models will
have the same 0-fermion sector, they will have different states at higher
fermion number\footnote{This is in addition to the more obvious observation
that one can consider two generalizations of the original Hamiltonian,
one in which it appears in the zero-fermion sector and one in which
it appear in the $N$-fermion sector.  This, however, is unrelated to
shape invariance, and merely involves switching $\psi_i$
and $\psi_i^\dagger$.}.

Thus, because of the $N$-body shape invariance,
we have the
two supersymmetric Hamiltonians, which have the same bosonic sector
but are otherwise distinct,
\beq
H_{S1} = \{A^\dagger_i(\alpha)\psi_i^\dagger,A_i(\alpha)\psi_i\}~,
\eeq
and
\beq
H_{S2}=\{A_i(\alpha+1)\psi_i^\dagger,
     A_i^\dagger(\alpha+1)\psi_i\}~.
\eeq

\section{Ending Remarks}

We have identified the algebraic identity that we expect underlies
the exact solubility of the Calogero and Calogero-Sutherland
models.  This identity is a natural 
$N$-body generalization of shape invariance, and we have explored
its implications for the analysis of these models, indicating in 
particular how it relates to integrability.  We have also seen how
this identity can be incorporated directly into our understanding
of the supersymmetric Calogero and Calogero-Sutherland models; indeed,
it appears that the supersymmetric context will be the natural arena
for using the generalized shape invaraince to solve the original
non-supersymmetric models.  Thus, the work in this paper should
provide a foundation for obtaining a simpler 
understanding of the 
solutions of Calogero 
and Calogero-Sutherland models, and a deeper understanding
more generally of exact solubility in the $N$-body case.
One particular point to mention is that
the eigenfunctions of the Calogero-Sutherland model
are deeply related to the so-called Jack polynomials of
the mathematics literature. These polynomials are defined in the space
of symmetric polynomials obeying certain conditions \cite{Mac}.
Although, they are quite well studied, mathematicians are
lacking an explicit representation. Recently, Lapointe and Vinet
presented such a formula \cite{LV}. Our proposal for
an extended shape invariance principle in higher than one dimensions
might prove useful to either derive this formula
in a compact and natural way or find an alternative simpler one.

\vspace{.5in}
\leftline{\Large\bf Acknowledgments}
D.S. thanks theory group at NIKHEF, where much of this work was
completed, for its hospitality.
D.S. was supported in part by the NSF through grant 
PHY-9509991.

\newpage
\appendix
\section{Theory of Many-Body Shape Invariant Potentials} 
\label{appendixA}

Here, we systematize the theory of $N$-body shape invariant potentials.
We develop the theory in a completely analogous way to the
1-dimensional theory so that the similarities and differencies
are apparent.

\subsection{Factorization of the Schr\"odinger Equation in the Many-Body Case}

Let us consider a potential $V_0(x_1,x_2,\dots,x_N)$
with ground state wavefunction 
$\Psi_0^{(0)}(x_1,\dots,x_N)$ and ground state energy
 $E_0^{(0)}$. As in the 2-particle case,
 without loss of generality,
we can assume that $E_0^{(0)}=0$. Now from the time-independent
 Schr\"odinger equation
\beql{schr1}
\label{schr1}
     H^{(0)}\,\Psi_0^{(0)}\=E_0^{(0)}\,\Psi_0^{(0)}\=0~,
\eeql
where
\beql{Nham0}
\label{Nham0}
     H^{(0)}\=-\sum_{i=1}^N \,\partial_i^2 \+ V_0(x_1,x_2,\dots,x_N)~,
\eeql
where $\partial_i\equiv \partial/\partial x_i$.

Obviously, the potential $V_0(x_1,x_2,\dots,x_N)$
is related to the ground state wavefunction
$\Psi_0^{(0)}$ by the equation: 
\beql{Npot0}
\label{Npot0}
    V_0(x)\= \sum_{i=1}^N \, 
   {\partial_i^2\Psi_0^{(0)}\over\Psi_0^{(0)}}~.
\eeql

In this case we introduce, the following creation and annihilation operators:
\beqnl{Ncran}
\label{Ncran1}
   A^\dagger_i &\eq&  -\partial_i -
       {\partial_i\Psi_0^{(0)}\over\Psi_0^{(0)}} ~,  \\
\label{cran2}
   A_i &\eq&  \phantom{-}\partial_i -
       {\partial_i\Psi_0^{(0)}\over\Psi_0^{(0)}}   ~.
\eeqnl

We clearly see
\beql{hamil0} 
\label{hamil0} 
  H^{(0)}\= \sum_{i=1}^N \,A^\dagger_i A_i ~.
\eeql

Finally, in the many-body case we define 
\beqnl{Nsuper}
\label{Nsuper}
   W_i &\eq& 
       -{\partial_i \Psi_0^{(0)}\over\Psi_0^{(0)}}  ~.
\eeqnl
In terms of the functions $W_i$, the creation and annihilation operators
take the form
\beqnl{NcranW}
\label{NcranW1}
   A^\dagger_i &\=&- \partial_i
       +W_i ~,  \\
\label{cranW2}
   A_i &\=& \partial_i
       +W_i ~.
\eeqnl

\subsection{A Special Class of Potentials}

A special class of potentials that has been discussed thoroughly in the 
literature is the class of potentials for which the ground state can be written
in the form
\beq
  \Psi_0^{(0)} \= \prod_{i=1}^N \prod_{j=i+1}^N\, 
  \psi_0^{(0)}(x_i-x_j)~.
\eeq

This demand leads to the condition
\beq
  W_i \= -  \sum_{j=1}^N\, {\partial_i\psi_0^{(0)}(x_i-x_j)\over
                                       \psi_0^{(0)}(x_i-x_j) }~.
\eeq
If $W(x)$ is the corresponding prepotential for the 2-body 
problem, we see that 
\beq
  W_i \=  \sum_{j=1}^N{}'\, W(x_i-x_j)~.
\eeq

Now, the previous choice of the wave function determines uniquely
the potential $V_0(x_1,x_2,\dots,x_N)$. 
Using relation \calle{Npot0}, we find that
\beqn
  V_0(x_1,x_2,\dots,x_N) &=&
     \sum_{j\ne i} \,
        {\partial_i^2\psi_0(x_i-x_j)\over \psi_0(x_i-x_j)}
    + \sum_{i\ne j\ne k\ne i}\,
    {\partial_i\psi_0(x_i-x_j)\over \psi_0(x_i-x_j)}
    \, {\partial_i\psi_0(x_i-x_k)\over \psi_0(x_i-x_k)}
   \nonumber \\
  &=&
     \sum_{j\ne i} \,
        v_0(x_i-x_j)
    + \sum_{i\ne j\ne k\ne i}\,
    W(x_i-x_j)
    \, W(x_i-x_k)~.
\eeqn

We notice that the last term consists of a sum of terms
$$
  W(a-b)W(a-c)+
  W(b-a)W(b-c)+
  W(c-a)W(c-b)
$$
Let $A=a-b$, $B=b-c$, and $C=c-a$ for which $A+B+C=0$.
Although we are not
restricting ourselves to shape invariant potentials at this point,
we make the following observation. For the usual shape invariant potentials,
except the Morse potential,
$W(x)$ is an odd function; so it is quite useful to  
restrict ourselves to odd super-potentials. Therefore, the expression 
written above has the form
$$
  -W(A) W(C)-W(A) W(B)-W(C) W(B)~.
$$
One now sees that direct 3-body  interactions are avoided if there is
a function $\tilde v_0(x)$ such that
\beq
\label{condition0}
 - W(A) W(C)-W(A) W(B)-W(C) W(B) \= \tilde v_0(A) +\tilde v_0(B)
  +\tilde v_0(C)~.
\eeq
In this case the potential $V_0$ is written
\beq
\label{eq:potential0}
 V_0(x_1,\dots,x_N) \= \sum_{j\ne i} \,
        v_0(x_i-x_j)
    +\sum_{j\ne i} \,
      \tilde  v_0(x_i-x_j)~.
\eeq

There are a few known solutions \cite{C2} to equation \calle{condition0}:

\begin{center}
\begin{tabular}{|c|c|c|c|}\hline
  $W(x)$          & $ v_0(x)$ & $\tilde v_0(x)$ & $\psi_0(x)$  \\ \hline
  $a x+{b\over x}$ & ${b(b+1)\over x^2} +a^2\, x^2 +2ba+a$  &
  $ab+{a^2\over 2}\, x^2$ & $|x|^b\,e^{ax^2/2}$  \\ \hline
  $a \,{\rm sgn}x$ & $a^2\,{\rm sgn}^2x+a\,\delta(x)$ &
  $-{a^2\over 3}$ &  $e^{a|x|}$   \\ \hline
  $a \cot x$  & ${a(a-1)\over\sin^2x} -a^2$ &
  $-{a^2\over 3}$ & $|\sin x|^a$    \\ \hline
  $a \coth x$  & ${a(a-1)\over\sinh^2x} +a^2$ &
  $-{a^2\over 3}$ & $|\sinh x|^a$    \\ \hline
  $a \zeta(x)-{\zeta(\omega)\over\omega}\, x$  & $W^2(x)-W'(x)$ &
  $1/2\, {\cal P}(x)-W^2(x)$  &  
  $|\theta_1(\pi x/2\omega\, |\, ir/2\omega)|$ 
  \\ \hline
\end{tabular}
\end{center}
In the above table, $\zeta(x)$ is the Weierstrass zeta function,
$\zeta'(x)=-{\cal P}(x)$, and  $\omega$ is the  half-period \cite{AS}.

\subsection{Shape Invariant Many-Body Potentials}

We now introduce the depence of the potential \calle{eq:potential0}
on possible parameters $\alpha_0$:
\beq
 V_0(x_1,\dots,x_N;\alpha_0) \= \sum_{i,j=1, j\ne i}^N \,
        v_0(x_i-x_j;\alpha_0)
    +\sum_{i,j=1, j\ne i}^N \,
      \tilde  v_0(x_i-x_j;\alpha_0)~,
\eeq
As we have discussed, the $N$-body hamiltonian is
written as
\beqn
   H^{(0)}(\alpha_0) = \sum_{i=1}^N\, A_i^\dagger(\alpha_0) A_i(\alpha_0)~.
\label{eq:ham0}
\eeqn
We notice now that the operators $A(\alpha)$ and $A^\dagger(\alpha)$
satisfy the commutation relations
\beqn
      \lb A_i(\alpha), A_j(\alpha') \rb
       &=& + W'(x_i-x_j;\alpha)-W'(x_j-x_i;\alpha')~,\\
      \lb A^\dagger_i(\alpha), A^\dagger_j(\alpha') \rb
       &=&
      -W'(x_i-x_j;\alpha)+W'(x_j-x_i;\alpha')~, \\
      \lb  A_i(\alpha), A^\dagger_j(\alpha') \rb
       &=&
      -W'(x_i-x_j;\alpha)-W'(x_j-x_i;\alpha')
       ~,
\eeqn
if $i\ne j$ and
\beqn
       \lb A_i(\alpha), A_i(\alpha') \rb  &=&
     -\sum_k{}' \, W'(x_i-x_k;\alpha)
     +\sum_k{}' \, W'(x_i-x_k;\alpha')~,\\
       \lb A^\dagger_i(\alpha), A^\dagger_i(\alpha') \rb &=&
     +\sum_k{}' \, W'(x_i-x_k;\alpha)
     -\sum_k{}' \, W'(x_i-x_k;\alpha')~,\\
      \lb A_i(\alpha),A^\dagger_i(\alpha') \rb &=&
      +\sum_k{}' \, W'(x_i-x_k;\alpha)
      +\sum_k{}' \, W'(x_i-x_k;\alpha')~ .
\eeqn
These commutation relations forbid us from repeating
the standard discussion of 1-dimensional
shape invariance (see section \ref{sec:2}), and thus
there is no automatic equivalence of the spectra for the Hamiltonian
\calle{eq:ham0} and
\beqn
   H^{(1)}(\alpha_0) = \sum_{i=1}^N\, A_i(\alpha_0) A_i^\dagger(\alpha_0)~.
\label{eq:ham1}
\eeqn
However, if
\beqn
    \sum_i \, A_i(\alpha_0) A_i^\dagger(\alpha_0) =
    \sum_i\, A_i^\dagger(\alpha_1) A_i(\alpha_1)
    + R(\alpha_1)~,
\eeqn
the ground states of the two hamiltonians are related by
\beq
     \Psi_0^{(1)}(\alpha_0)
     \propto
     \Psi_0^{(0)}(\alpha_1)~.
\eeq
Furthemore, the ground state of $H^{(1)}$ has energy $R(\alpha_1)$.

Obviously, one can  repeat the precedure,
exactly as in the 1-dimensional case,  and obtain the series
of Hamiltonians
\beq
    H^{(n)}(\alpha_0) =
   H^{(0)}(\alpha_n) + \sum_{k=1}^n \, R(\alpha_k)~.
\eeq
The corresponding ground states are given by
\beq
     \Psi_0^{(n)}(\alpha_0)
     \propto
     \Psi_0^{(0)}(\alpha_n)~,
\eeq
and have energies
\beq
   E^{(n)}_0 =  \sum_{k=1}^n \, R(\alpha_k)~.
\eeq
Unfortunately, there is no immediate relationship between these
states and the spectrum of the original model.


\bye